\documentclass[conference]{IEEEtran}
\IEEEoverridecommandlockouts
\usepackage{cite}
\usepackage{amsmath,amssymb,amsfonts}
\usepackage{algorithm}
\usepackage{graphicx}
\usepackage{textcomp}
\usepackage{subcaption}
\usepackage{url}
\usepackage{xcolor}
\usepackage{algpseudocode}
\usepackage[margin=0.75in]{geometry}
\usepackage{times}
\usepackage{hyperref}
\def\BibTeX{{\rm B\kern-.05em{\sc i\kern-.025em b}\kern-.08em
    T\kern-.1667em\lower.7ex\hbox{E}\kern-.125emX}}

\DeclareRobustCommand*{\IEEEauthorrefmark}[1]{%
    \raisebox{0pt}[0pt][0pt]{\textsuperscript{\footnotesize\ensuremath{#1}}}}

\begin{document}
\title{Adaptive Configuration Selection for Multi-Model Inference Pipelines in Edge Computing
}

\author{
\IEEEauthorblockN{
Jinhao Sheng\IEEEauthorrefmark{1},
Zhiqing Tang\IEEEauthorrefmark{1},
Jianxiong Guo\IEEEauthorrefmark{1,2}, and
Tian Wang\IEEEauthorrefmark{1}}
\IEEEauthorblockA{\IEEEauthorrefmark{1}Institute of Artificial Intelligence and Future Networks, Beijing Normal University, Zhuhai, China}
\IEEEauthorblockA{\IEEEauthorrefmark{2}Guangdong Key Lab of AI \& Multi-Modal Data Processing, BNU-HKBU United International College, Zhuhai, China}
\IEEEauthorblockA{jhsheng@cmu.edu.cn, \{zhiqingtang, jianxiongguo, tianwang\}@bnu.edu.cn}
\thanks{This work is supported in part by the National Natural Science Foundation of China (NSFC) under Grant 62302048 and Grant 62272050. \textit{(Corresponding author: Zhiqing Tang.)}}
}

\maketitle

\begin{abstract}

The growing demand for real-time processing tasks is driving the need for multi-model inference pipelines on edge devices. However, cost-effectively deploying these pipelines while optimizing Quality of Service (QoS) and costs poses significant challenges. Existing solutions often neglect device resource constraints, focusing mainly on inference accuracy and cost efficiency. To address this, we develop a framework for configuring multi-model inference pipelines. Specifically: 1) We model the decision-making problem by considering the pipeline's QoS, costs, and device resource limitations. 2) We create a feature extraction module using residual networks and a load prediction model based on Long Short-Term Memory (LSTM) to gather comprehensive node and pipeline status information. Then, we implement a Reinforcement Learning (RL) algorithm based on policy gradients for online configuration decisions. 3) Experiments conducted in a real Kubernetes cluster show that our approach significantly improve QoS while reducing costs and shorten decision-making time for complex pipelines compared to baseline algorithms.
\end{abstract}
\section{Introduction}


The rapid advancement of machine learning and large-scale deep learning models has led to a surge in demand for applications that can process tasks instantly. Applications such as Internet of Things (IoT) \cite{3}, autonomous driving \cite{1}, and healthcare \cite{47} depend on complex model inference pipelines for stability and reliability in challenging situations. Many companies currently deploy their machine learning model inference pipelines on cloud platforms \cite{5}, but high latency and heavy reliance on network bandwidth often hinder real-time processing capabilities. Edge computing \cite{4} addresses this issue by moving data processing, storage, and application services from centralized data centers to the network edge, significantly reducing latency and making it ideal for real-time processing scenarios \cite{42} \cite{51}. Consequently, deploying multi-model inference pipelines on edge devices is increasingly essential to meet these demands.

Deploying multi-model inference pipelines on edge devices can significantly reduce issues like latency, but finding a cost-effective way to implement these pipelines remains a major issue. Multi-model inference pipelines, which form a chain of machine learning models, raise several concerns regarding Quality of Services (QoS) \cite{13} and cost reduction. Unlike optimizations that focus on a single stage, optimizing an end-to-end inference pipeline requires a comprehensive approach that takes into account configuration adjustments across multiple steps. Previous works \cite{16,26,46} have proposed solutions like efficient autoscaling, batching, and pipeline scheduling. These approaches address the aforementioned issues and also considered the dynamic nature of machine learning workloads. Existing methods often overlook critical factors such as inference accuracy and device resource constraints. Our work thoroughly considers these factors and provides a detailed model for configuring multi-model inference pipelines.


Configuring multi-model inference pipelines online to improve QoS and reduce costs holds great promise. However, several challenges need to be addressed. \textit{The first challenge is how to effectively extract features from the dynamic load environment of these pipelines.} While existing methods like Fast, Accurate Autoscaling (FA2) \cite{17} and Inference Pipeline Adaptation (IPA) system \cite{15} offer some solutions, they often fail to comprehensively consider aspects like pipeline accuracy, cost control, and device resources. To tackle this issue, we propose a feature extraction module based on residual networks \cite{36}, which can deeply mine the state information of nodes and pipelines. Additionally, given the dynamic nature of workloads, accurately predicting incoming workloads is vital for adjusting the configurations of multi-model inference pipelines. To meet this need, we design a load prediction model based on Long Short-Term Memory (LSTM) \cite{35}. The combination of the feature extraction module and the load prediction model provides robust real-time data support for configuration decisions.


Then, \textit{the second challenge is how to effectively coordinate the decisions across multiple stages of the inference pipeline, including choices about replication factors, batch sizes, and model variations.} In the pipeline, the model selection at the earlier stages directly affects the best choices for the downstream models. The performance of the earlier models directly impacts the end-to-end latency and accuracy, which in turn limits the available options for downstream models. Although there are some heuristic algorithms like autoscaling \cite{8} and model switching \cite{9} trying to address this issue, these methods still fall short in coordinating decision-making for the various stages of the inference pipeline. To tackle this problem, we propose an Online Pipeline Decision (OPD) algorithm based on policy gradients \cite{7} that optimizes decision-making and flexibly adapts to changing workloads.


In this paper, we first model the online configuration selection problem for multi-model inference pipelines as a Markov Decision Process (MDP), aiming to maximize QoS while minimizing costs. Based on this, we propose a Reinforcement Learning (RL) algorithm using policy gradients for real-time configuration adjustments. To implement and evaluate this algorithm, we deploy multi-model inference pipelines in a real Kubernetes \cite{18} cluster. Specifically, we use Selcon Core \cite{20} and MLServer \cite{21} to set up the pipeline, leveraging Prometheus \cite{19} to monitor node status and track load information, which provides the necessary state data for the RL Agent's decisions. We then adjust the configuration via the Kubernetes Python API to implement these decisions. The experimental results show that our algorithm outperforms existing baseline algorithms in enhancing QoS and reducing costs, while also significantly shortening decision-making time.


To sum up, the contributions of this paper are as follows:

\begin{itemize}
\item We model the online configuration decision problem in a multi-model inference pipeline as a MDP for the first time. Our goal is to maximize QoS while minimizing costs.
\item We design a feature extraction module to extract information from nodes and the pipeline and an LSTM-based network to predict upcoming loads. We then present the OPD algorithm, enabling dynamic adaptation and optimization of decision-making.
\item We validate the OPD algorithm's effectiveness in a real Kubernetes cluster, demonstrating improved QoS, reduced costs, and significantly faster decision-making, surpassing all baseline algorithms.
\end{itemize}


\section{Related Work}
\label{section: related work}

\textbf{Multi-stage Inference Serving:} As the complexity and scale of machine learning applications grow, selecting configurations for model inference serving has garnered attention, especially in multi-stage inference services \cite{16,17,26,25}. Recent studies have investigated methods to improve inference performance in these systems. Crankshaw \textit{et al.} \cite{26} use a heuristic approach to optimize pipeline configurations based on Service Level Objectives (SLOs), effectively reducing end-to-end latency. Razavi \textit{et al.} \cite{17} automate scaling with target graph transformations and a dynamic programming algorithm. Ghafouri \textit{et al.} \cite{15} present IPA, a system that balances accuracy and cost objectives according to designer preferences. Salmani \textit{et al.} \cite{9} propose a strategy for dynamically switching between lightweight and heavy models based on workload, adapting to high-load conditions with models that have higher throughput but lower accuracy. Gunasekaran \textit{et al.} \cite{31} offer a cost-saving ensemble learning method that meets prior latency and accuracy requirements, despite being costly. However, existing methods often overlook model switching, cost control, QoS, or resource constraints in edge devices.


\textbf{RL in Edge Computing:} In recent years, reinforcement learning (RL) algorithms have demonstrated outstanding performance in problems requiring long-term decision-making and strategy optimization. Many studies now apply RL methods to challenges in edge computing. Chen \textit{et al.} \cite{12} model data and service placement as MDP, using RL to achieve optimal placement decisions. Wang \textit{et al.} \cite{40} address service migration as observable MDP, implementing RL strategies for optimal migration. RL applications also extend to container migration \cite{43}, traffic control \cite{42}, task offloading \cite{41}, and container scheduling \cite{44}. Building on this research, we analyze the feasibility of using RL for configuring a multi-model inference pipeline in edge computing.
\begin{figure}[!ht]
 \centering
 \includegraphics[width=1\linewidth]{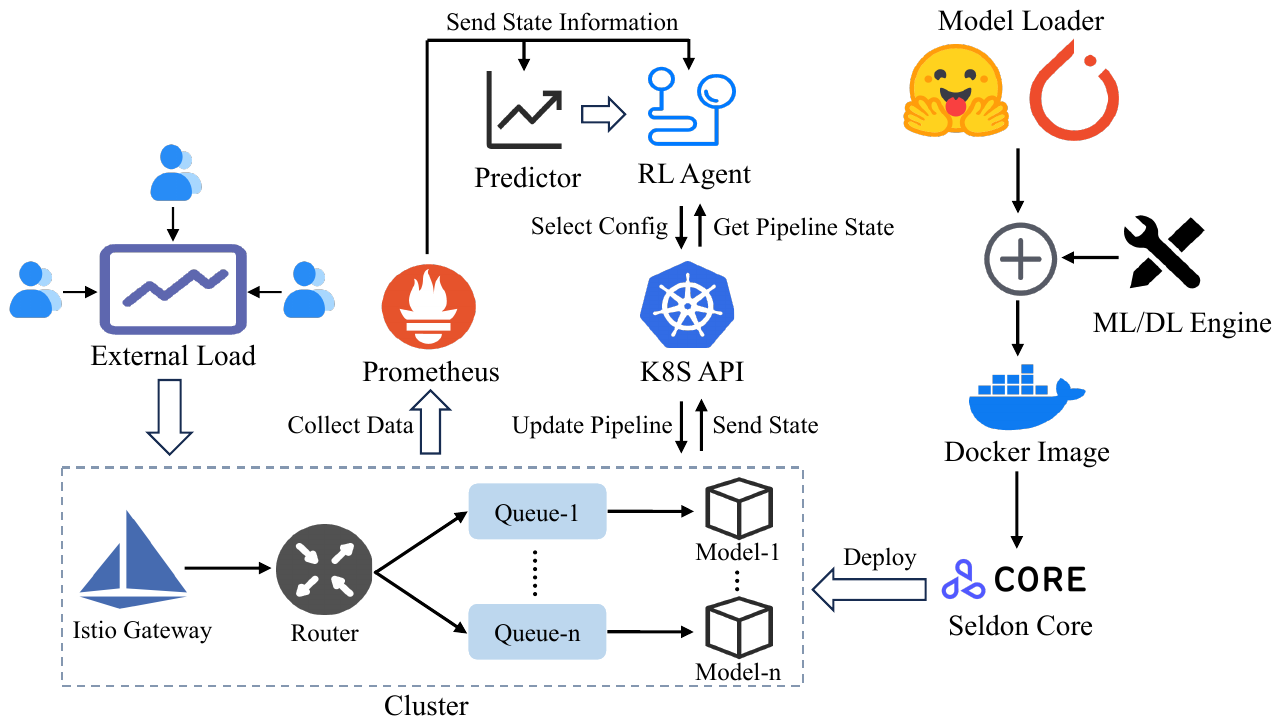}
 \centering
 \caption{System Overview.}
 \label{fig: system}
\end{figure}
\section{System Design And Problem Formulation}
\label{section: system design and problem formulation}
\subsection{System Design}
This section provides an overview of the system's main components, as illustrated in Fig. \ref{fig: system}. The system consists of four core components: model loading, monitoring, pipeline system, and RL agent.

\textbf{Model Loading:} Users submit models for each pipeline stage, ensuring a balance of accuracy, latency, and resource usage. Model variants can be created using optimizers like TensorRT \cite{49} and ONNX graph optimization \cite{48}, employing various quantization levels and Neural Architectural Search methods. Additionally, we store these model variants in an object storage service to minimize container creation time.

\textbf{Monitoring:} The monitoring daemon utilizes the highly available Prometheus \cite{19} time-series database to periodically track the incoming load to the system and the pipeline.

\textbf{Pipeline System:} Once model loading is complete, the inference pipeline becomes fully operational, allowing users to send requests that traverse the models and return inference results. A centralized load balancer distributes these requests across the pipeline stages, each supported by a centralized queue to ensure predictable behavior and efficient latency modeling. GRPC \cite{23} facilitates communication between the pipeline stages. Load balancing among containers within the same stage is managed by Istio sidecar containers \cite{22}. Each model container is deployed using Docker, sourced from a forked MLServer \cite{21} and Seldon Core version, to implement gRPC web servers on Kubernetes. Detailed deployment steps are provided in Section \ref{section: system implementation}.

\textbf{RL Agent:} The RL agent periodically performs the following tasks: (1) retrieves incoming load and node information from the monitoring daemon, (2) uses a predictor to forecast the next reference load based on historical data, (3) obtains current pipeline configuration details via the Kubernetes Python API, and (4) selects and applies a configuration based on this information.

\begin{figure*}[htbp]
    \centering
    \includegraphics[width=\textwidth]{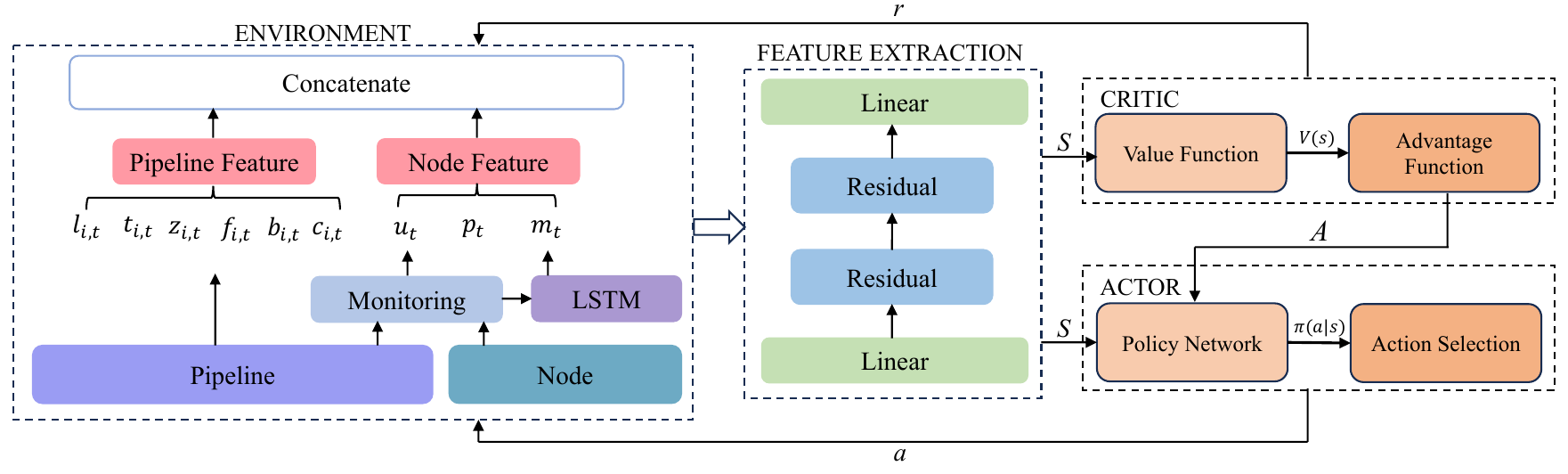}
    \caption{Algorithm Overview.}
    \label{fig: algorithm}
\end{figure*}

\subsection{Definition of Metrics Over Pipeline}

We examine a multi-model inference pipeline that includes key elements: task, accuracy, cost, and QoS \cite{13}. These components define the system's operational and performance characteristics.

\textbf{Task:} The multi-model inference pipeline comprises a set of tasks denoted as $n$ $\in$ $\textbf{N}=\left \{n_1,n_2,\ldots,n_{\left |\textbf{N}  \right | }\right \}$, where $\left|\cdot\right|$ represents the number of tasks, such that $\left|\textbf{N}\right|$ indicates the total number of tasks. Each task has an associated set of model variants $\textbf{Z} = \left \{z_1,z_2,\ldots,z_{\left |\textbf{Z}  \right | }\right \}$, which may differ in quantity. The accuracy and cost of task $n$ using model variant $z_i$ are denoted as $v_n(z_i)$ and $c_n(z_i)$, respectively. Additionally, the replication factor for task $n$ is $f_n$, and the batch size is $b_n$. Throughput and latency for task $n$ are represented by $t_n$ and $l_n$, respectively.

\textbf{Accuracy:} To assess the accuracy of a multi-model inference pipeline, end-to-end accuracy is typically calculated using labeled data. However, this approach may not adequately capture the complexity of the semantic stages or the variability among models at each stage. To address this, we have adopted a heuristic approach inspired by \cite{15}, focusing on linear inference pipelines with a single input and output connected by a series of models. The accuracy of each model is pre-calculated offline, allowing us to determine the overall accuracy of the pipeline by:
\begin{equation}
\begin{matrix}
    V = \sum_{n \in N, z_i \in Z}v_n(z_i),
\end{matrix}
\end{equation}
where $v_n(z_i)$ denotes the accuracy of the model variant $z_i$ used in task $n$.\\
\indent \textbf{Cost:} In our multi-model inference pipeline, cost refers to the CPU cores allocated to model variants. In Kubernetes, CPU resources are assigned by indicating the number of cores, which helps evaluate the pipeline's economic efficiency. The total cost of the pipeline can be calculated as:
\begin{equation}
\begin{matrix}
    C = \sum_{n\in N, z_i \in Z}f_n\times c_n(z_i),
\end{matrix}
\end{equation}
where $f_n$ denotes the number of replicas for task $n$, and $c_n(z_i)$ indicates the resource usage of model variant $z_i$ for task $n$.

\textbf{QoS:} We have established specific QoS metrics for our multi-model inference pipelines, highlighting their unique attributes: accuracy, throughput, latency, and extra load as key performance indicators. We calculate QoS as follows:
\begin{equation}
    Q = \left\{
    \begin{array}{ll}
      \alpha \times V + \beta \times T - L - \gamma \times E, & \text{if } e \geq 0 \\
      \alpha \times V + \beta \times T - L - \delta \times (-E), & \text{if } e < 0
    \end{array} \right. 
    \label{eq: QoS}
\end{equation}
where $\alpha$, $\beta$, $\gamma$, and $\delta$ are weighting parameters; $T$ is the pipeline's throughput, defined as the minimum throughput across all tasks; $L$ is the pipeline's latency, which is the sum of $l_n$; $E$ represents excess load, indicating the unprocessed portion of demand due to resource constraints. Positive $E$ values signal unmet demand, while negative values indicate spare capacity. This definition of QoS enables us to assess the inference pipeline's performance from multiple dimensions.

\subsection{Problem Formulation}
Our objective is to improve QoS while simultaneously reducing costs. We aim to identify the optimal strategy that balances QoS and cost efficiency, defined as follows:
\begin{equation}
\begin{aligned}
\max T &= Q - \lambda \times C, \\
\text{s.t.} \quad & \text{Eqs. } (1),(2),(3),\\
& 0 < i \le \left|\textbf{Z}\right|,\forall z_i \in Z,\\
& 0 < f_n \le F_{max} ,\forall n \in N, \\
& 0 < b_n \le B_{max} ,\forall n \in N, \\
& \sum_{n \in N, z_i \in Z}w_n(z_i) \times f_n \leq W_{\text{max}},
\end{aligned}
\end{equation}
where $F_{max}$ is the maximum allowed replication factor, $B_{max}$ is the maximum permissible batch size in a pipeline, $w_n(z_i)$ represents the resource consumption of model variant $z_i$ for task $n$, and $W_{max}$ is the total resource capacity of the device.

Configuring multi-model inference pipelines is an NP-hard decision problem, meaning that finding or verifying an optimal solution is computationally difficult. Common approaches include heuristic algorithms, approximation techniques, and decomposition methods, each with specific limitations. RL algorithms present a promising alternative, particularly in complex decision environments where optimal strategies are not obvious. The configuration process in multi-model inference pipelines is dynamic, with each decision affecting current and future states. This sequential decision-making aligns well with MDP, making it suitable to model the problem as an MDP.

\section{Our Algorithms}
\label{section: algorithms}

\subsection{Workload Prediction}

We have developed an LSTM \cite{35} model to predict the maximum workload for the next 20 seconds based on a time series of loads per second collected over the past 2 minutes. The model architecture includes a 25-unit LSTM layer followed by a one-unit dense output layer.

\subsection{RL Settings}

\textbf{State:} The state $s_t$ comprises the node state and the pipeline state. The node state encompasses available resources, incoming loads, and predicted loads, while the pipeline state records each task's current model index, number of replicas, batch size, costs, latency, and throughput performance. Thus, the state $s_t$ is defined as:
\begin{equation}
\begin{aligned}
s_{t} = &[(u_{t},p_{t},m_{t},l_{1,t},t_{1,t},z_{1,t},f_{1,t},b_{1,t},c_{1,t}),\\
&\hspace{2.5cm} \ldots,\\
&(u_{t},p_{t},m_{t},l_{n,t},t_{n,t},z_{n,t},f_{n,t},b_{n,t},c_{n,t})].
\label{eq: state}
\end{aligned}
\end{equation}

\indent \textbf{Action:} The action $a_t$ encompasses the configuration for each pipeline task, including the model variant index, number of replicas, and batch size. Thus, $a_t$ can be defined as:
\begin{equation}
    a_{t} = [(z_{1,t},f_{1,t},b_{1,t}),\ldots,(z_{n,t},f_{n,t},b_{n,t})].
\end{equation}
When the task changes, the action space must be modified to match the configuration options of each pipeline task.

\textbf{Reward:} Given that various workloads require distinct configurations, we have optimized the reward function to improve the stability and effectiveness of policy gradient algorithms while achieving our desired objectives. The reward function is defined as follows:
\begin{equation}
    {r_{t}} = Q - \beta \times C - \gamma \times B,
\end{equation}
where the parameter $\beta$ represents the weight assigned to the cost, $\gamma$ is the penalty coefficient for the batch sizes, and $B$ denotes the maximum batch sizes $b_n$ for each task in a pipeline. $\gamma$ is employed to prevent excessively large batch sizes, ensuring that configurations enhance QoS while maintaining a reasonable scale.

\begin{algorithm}[t]
\caption{The OPD Algorithm}
\label{algorithm: alg1}
\textbf{Input}: $L$, $P$, $M$ \\
\textbf{Output}: $H$
\begin{algorithmic}
\State Deploy Pipeline in Cluster;
\State Load the OPD Agent;
\State Reset environment and get state $s_{0}$;
\State Incoming workloads;
\For{$t = 1,2,\ldots$}
    \State Predict the coming workload;
    \State Get the state $s_{t}$ by Eq. \eqref{eq: state};
    \State Select $a_{t}$ according to the OPD agent;
    \State Calculate the decision time $d_t$;
    \State Apply the action $a_{t}$ and convert it into a configuration
    \State that can be applied to the cluster;
    \State Apply new configuration by Kubernetes Python API;
    \State Calculate reward $r_{t}$;
    \State Get the next state $s_{t+1}$;
\EndFor
\State Compute the cumulative decision time: $H = \sum d_{t}$;
\end{algorithmic}
\end{algorithm}
\subsection{Online Pipeline Decision}
In this section, we introduce the Online Pipeline Decision (OPD) algorithm, detailing the feature extraction process and the training methodology.

\textbf{Overview:} The OPD algorithm framework is shown in Fig. \ref{fig: algorithm}. The system retrieves the resource status, load conditions, and task configurations from the environment. This information is fed into feature extraction and processed by the policy network to make decisions and calculate decision-time, as detailed in Algorithm \ref{algorithm: alg1}. Rewards are generated based on the actions taken, and the policy network alongside the value functions is updated using policy gradient algorithms \cite{7}.

\textbf{Feature Extraction:} The feature consists of two main components: node features and pipeline features, which are combined into a unified feature vector. Raw data from node and pipeline states undergoes processing through a fully connected layer to reduce dimensionality to a manageable intermediate form. This processed data is then refined through several residual blocks \cite{36}.

\textbf{Training:} The OPD algorithm focuses on policy optimization, where a policy, typically denoted as $\pi$, guides the agent's action choices, expressed as $a(t) \sim \pi(\cdot|s(t))$. In multi-model inference scenarios, the probability of the decision process is defined as:
\begin{equation}
\begin{matrix}
    P(\gamma |\pi ) = \rho_{0} (s_{0})\prod_{t=0}^{T-1}P(s_{t+1}|s_{t},a_{t})\pi (a_{t}|s_{t}),
\end{matrix}
\end{equation}
where $\rho_{0} (s_{0})$ is the start-state distribution.

In policy gradient algorithms, the most commonly used gradient estimator $\hat{g}$ has the form of:
\begin{equation}
\begin{matrix}
\hat{g} = \hat{\mathbb{E}}_{t}[\bigtriangledown _{\theta} log \pi_{\theta}(a_{t}|s_{t})\hat{A}_{t}],
\end{matrix}
\end{equation}
which is obtained by differentiating the loss function $L^{PG}(\theta)$ of policy gradient:
\begin{equation}
\begin{matrix}
    L^{PG}(\theta) = \hat{\mathbb{E}}_{t}[log \pi_{\theta}(a_{t}|s_{t})\hat{A}_{t}].
\end{matrix}
\end{equation}
However, traditional policy gradient methods can be inefficient and lead to disruptive updates, resulting in instability. To mitigate these issues, Proximal Policy Optimization (PPO) \cite{33} uses a clipped surrogate objective:
\begin{equation}
    L_t(\theta) = \hat{\mathbb{E}_t}[L^{CLIP}_t(\theta) - c_1L^{VF}_t(\theta) + c_2S[\pi_\theta](s_t)],
    \label{eq: loss}
\end{equation}
where $c_1$, $c_2$ are coefficients, and $S$ denotes an entropy bonus, and $L_t^{VF}$ is a squared-error loss. The $L_t^{CLIP}(\theta)$ is denoted as:
\begin{equation}
L_t^{CLIP}(\theta) = \hat{\mathbb{E}}_{t}[\min(r_{t}(\theta)\hat{A}_{t},clip(r_{t}(\theta),1-\epsilon ,1+\epsilon)\hat{A}_{t}],
\end{equation}
where $r_{t}(\theta) = \frac{\pi_{\theta}(a_{t}|s_{t})}{\pi_{old}(a_{t}|s_{t})}$ is the policy ratio, $\hat{A}_{t}$ is the estimated advantage function, and $\epsilon$ is a small constant that limits the range of $r_{t}(\theta)$ to prevent training instabilities from large updates.

To tackle the issues of sparse training data and slow initial convergence, we have integrated expert guidance into our training. The training process for the OPD algorithm is detailed in Algorithm \ref{algorithm: alg2}.

\begin{algorithm}[t]
\caption{Training of the OPD Algorithm}
\label{algorithm: alg2}
\textbf{Input:} $P$, $L$\\
\textbf{Output:} $a_{t}$
\begin{algorithmic}
\State Initialize policy network $\pi_{\theta}$;
\State Initialize replay memory $\textbf{D} = \emptyset$;
\State Initialize expert optimizer as $expert\_model$;
\State Define expert\_frequency as $f$;
\For{$e \leftarrow$ 1,2, \ldots}
    \State Reset the environment and obtain the initial state $s_0$;
    \While{not completed processing of $L$}
    \If{$e \% f == 0$}
    \State $a_t \leftarrow$ action from expert\_model given $s_t$;
    \Else
    \State $a_t \leftarrow$ action from policy $\pi_{\theta}$ given $s_t$;
    \EndIf
    \State Execute action $a_t$ in the environment;
    \State Observe new state $s_{t+1}$ and calculate reward $r_t$;
    \State Store $\{s_t,a_t,r_t\}$ in $\textbf{D}$ and compute target value 
    \State $\hat{V}_{\pi}(s_t)$;
    \EndWhile
    \For{$t = 1,2,\ldots,T$}
    \State Compute $\hat{A_1},\ldots,\hat{A_T}$;
    \State Compute loss $L(\theta)$ by Eq. \eqref{eq: loss};
    \State Optimize the network by mini-batch SGD with 
    \State $Adam$ optimizer;
    \EndFor
    \State Update weights from time to time $\pi_{old}$ $\leftarrow$ $\pi_{\theta}$;
\EndFor   
\end{algorithmic}
\end{algorithm}
\subsection{Computational Complexity Analysis}

The OPD algorithm consists of four components: state observation, action selection, reward calculation, and network update. Below is an analysis of the computational complexity for each part. First, the state is defined in Eq. \eqref{eq: state}, with a complexity of $O(\left | N \right |)$. Second, the complexity of action selection for all pipeline tasks is also $O(\left | N \right |)$. The complexity of reward calculation remains constant regardless of the number of tasks, resulting in $O(1)$. Finally, feature extraction occurs through fully connected layers within residual blocks, where each layer spans $L$ layers and contains $G$ neurons. The complexity for this section is $O(\left | N \right |\times G + L \times G^{2})$.

The total experience used for each network update is constant, while the ratio of expert to agent experience varies. Other environmental operations have little effect on the computational complexity. Thus, the complexity of the OPD algorithm is expressed as $O(|N| + |N| \times G + L \times G^2)$.

\begin{figure}[!h]
 \centering
 \includegraphics[width=1\linewidth]{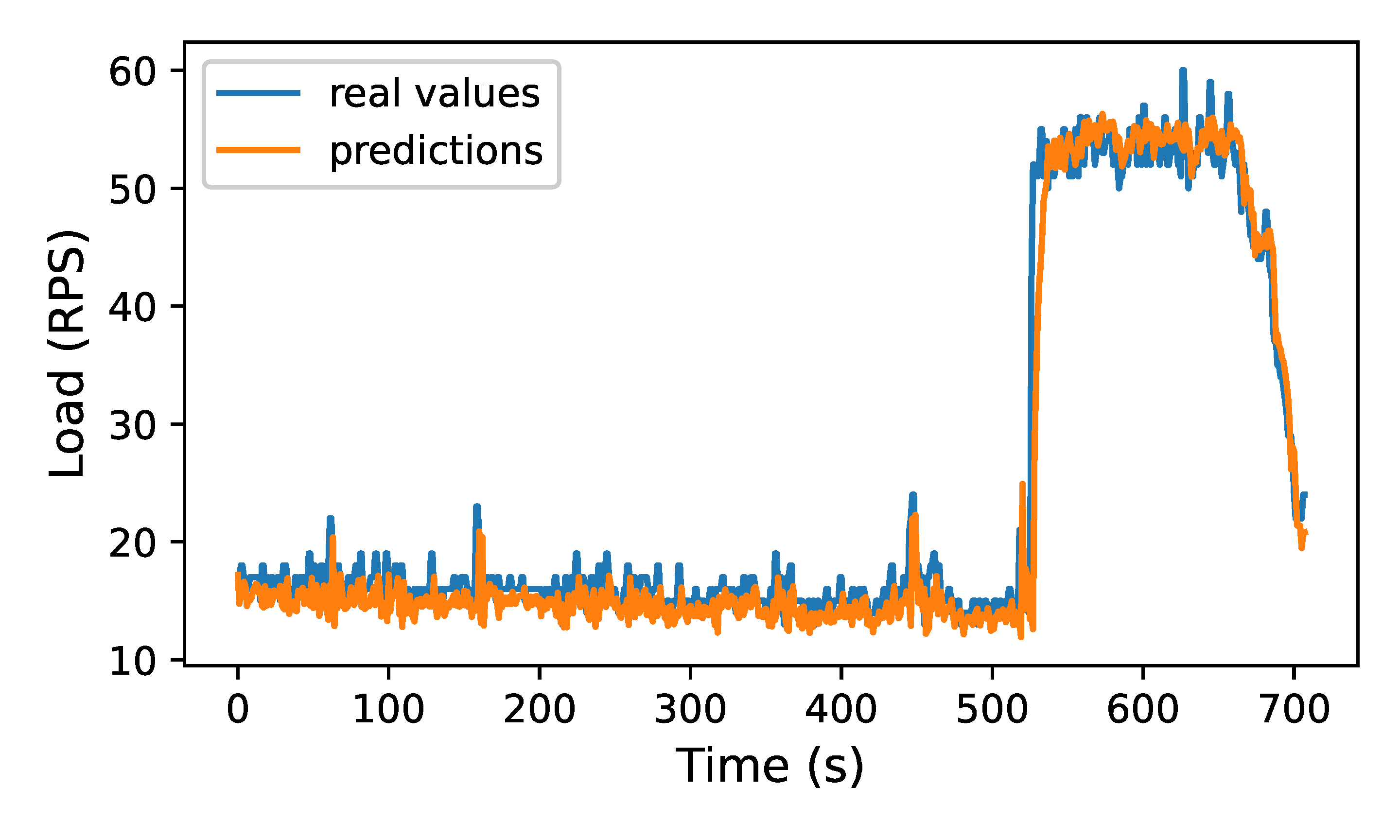}
 \centering
 \caption{Lstm Prediction.}
 \label{fig: lstm}
\end{figure}

\section{System Implementation}
\label{section: system implementation}

A multi-model inference pipeline is deployed on a Kubernetes cluster, a popular open-source platform for orchestrating containerized applications.

\subsection{Implementation Issues}

\textbf{Network Communication:} The system processes inference requests sequentially across multiple models, which can introduce latency. To mitigate this, it uses gRPC \cite{23} for inter-stage communication, optimizing data transfer efficiency and reducing delays.

\textbf{Multi-Model Dependency Management:} Different models may need various library versions or runtime environments, leading to dependency conflicts. Deploying models in Docker containers isolates these environments, preventing version conflicts and ensuring stable system operation.

\textbf{Resource Management and Scheduling:} A critical challenge during high-load scenarios is ensuring efficient resource utilization among multiple containers at the same stage. We implement Istio sidecar containers \cite{22} for load balancing and employ our OPD algorithm to dynamically adjust pipeline configurations based on available resources, maximizing resource usage and preventing any instance from being resource-constrained.

\subsection{Deployment of Multi-Model Inference Pipelines}

Machine learning models can be encapsulated using MLServer and Seldon Core tools to create Docker images. These images are deployed on a Kubernetes cluster via a ``SeldonDeployment" YAML file, which outlines deployment details such as resource requirements, replica counts, exposed ports, and other relevant information. 

In an Istio-supported environment, the multi-model inference pipeline is managed through Istio's routing rules, utilizing ``Virtual Service" and ``Destination Rule" to direct and optimize data flow between models, thereby enhancing the microservices' security, reliability, and efficiency.

Once deployed, Seldon Core's management interface and APIs allow for monitoring, management, and optimization of the model's operational status. For detailed performance monitoring and log analysis, integration with Prometheus and Grafana is also possible.

\begin{figure}[!h]
    \centering
    \begin{subfigure}[!h]{0.5\textwidth}
        \centering
        \includegraphics[width=\textwidth]{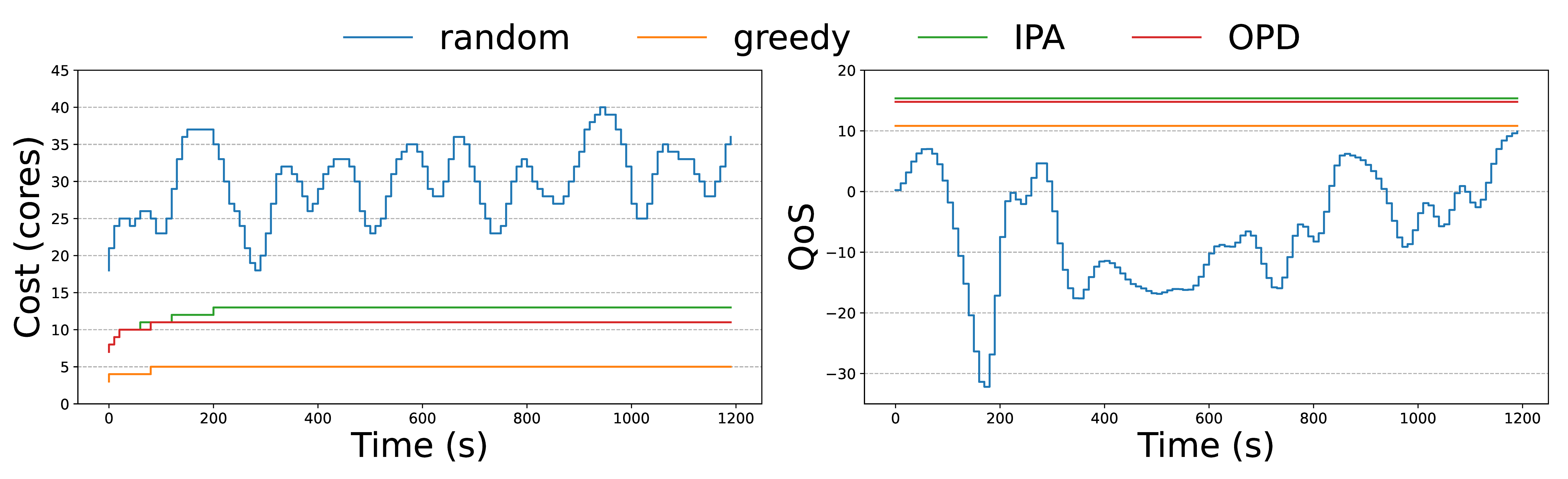}
        \caption{steady low workload}
        \label{fig: time_low}
    \end{subfigure}
    \begin{subfigure}[!h]{0.5\textwidth}
        \centering
        \includegraphics[width=\textwidth]{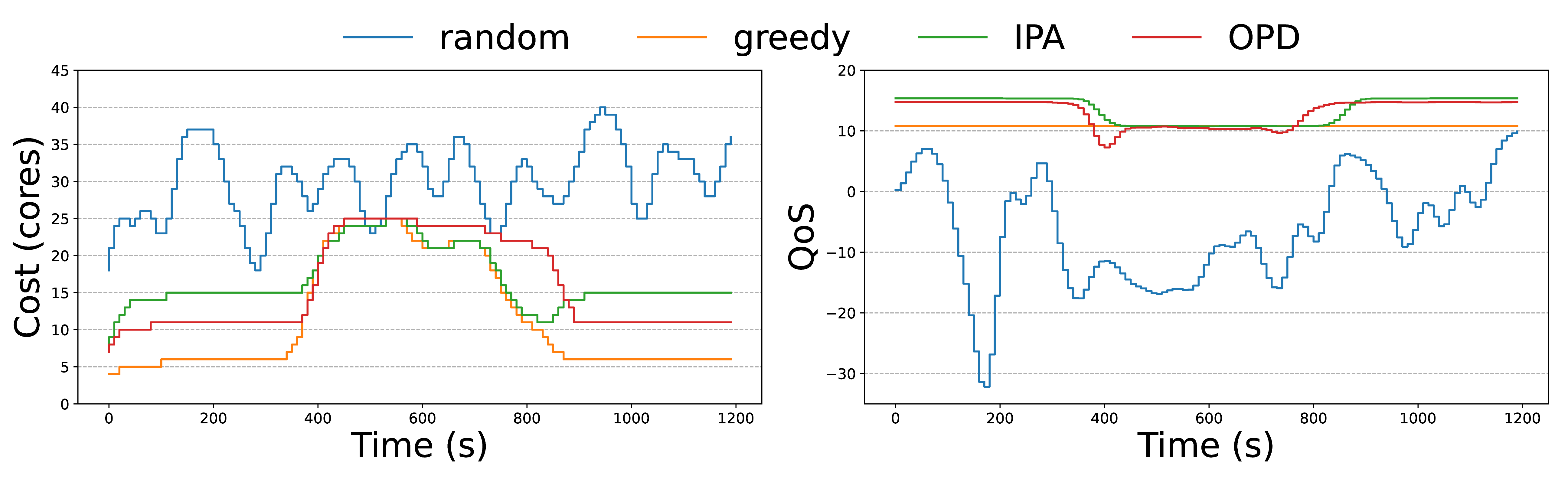}
        \caption{fluctuating workload}
        \label{fig: time_flu}
    \end{subfigure}
    \begin{subfigure}[!h]{0.5\textwidth}
        \centering
        \includegraphics[width=\textwidth]{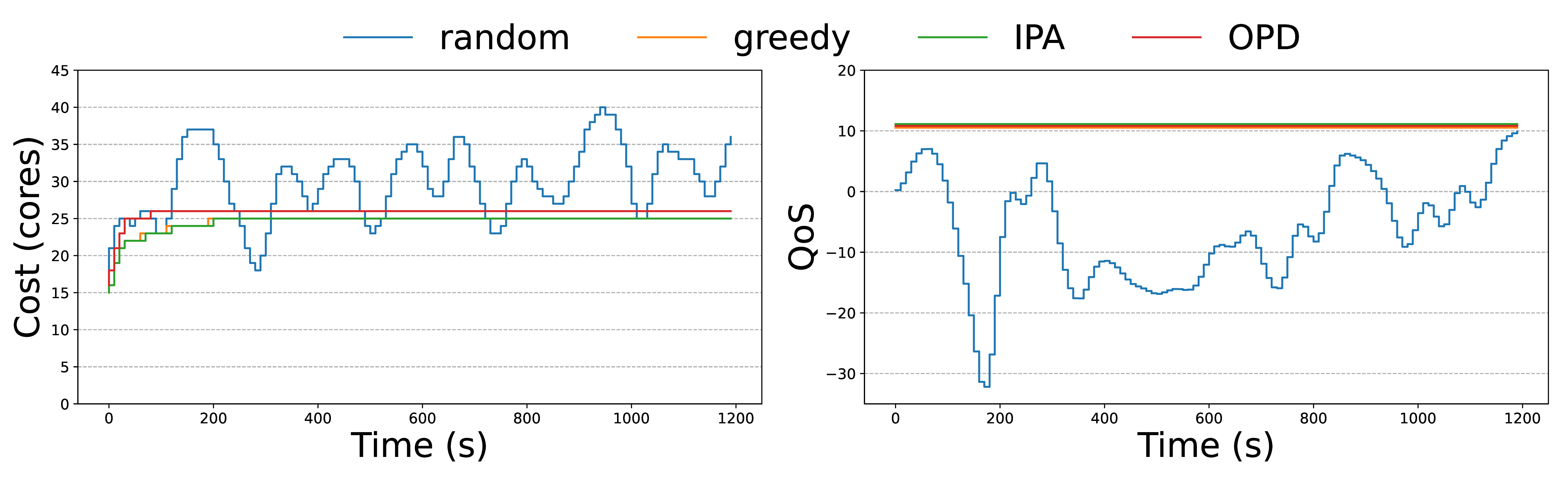}
        \caption{steady high workload}
        \label{fig: time_high}
    \end{subfigure}
    \caption{Temporal analysis under different workloads.}
    \label{fig: time}
\end{figure}

\section{Performance Evaluation}
\label{section: performance evaluation}
\subsection{Experimental Settings}

\textbf{Hardware:} We evaluate the OPD algorithm using three physical machines, each equipped with an Intel i9-10900K 10-core CPU, 32 GB of RAM, and an NVIDIA GeForce RTX 2070 Super graphics card.

\textbf{Predictor:} The LSTM module has been trained to predict workloads in under 50 milliseconds. As shown in Fig. \ref{fig: lstm}, the model demonstrates high accuracy in testing, achieving a Symmetric Mean Absolute Percentage Error (SMAPE) \cite{36} of only 6\%, comparable to predictors in similar systems.

\textbf{Pipelines:} Previous studies frequently relied on customized pipelines, which posed challenges in production settings. In contrast, our pipelines prioritize flexibility and versatility, incorporating insights from earlier research and industrial applications, as outlined in \cite{15}. Each task within our pipeline is supported by a selection of models tailored to specific requirements.

\textbf{Baselines:} Several baselines are conducted for performance comparison, detailed as follows.
\begin{enumerate}
    \item \textbf{Random}. The algorithm randomly selects configurations for each pipeline task, promoting diversity and exploration in selection.
    \item \textbf{Greedy}. A greedy algorithm chooses the configuration for each pipeline task to minimize costs while adhering to available resource constraints.
    \item \textbf{IPA} \cite{15}. This algorithm uses a solver to select configurations for each task in a pipeline but ignores resource constraints. We have enhanced the algorithm to factor in resource availability during configuration selection.
\end{enumerate}

\begin{figure}[!h]
    \centering
    \begin{subfigure}[!h]{0.5\textwidth}
        \centering
        \includegraphics[width=\textwidth]{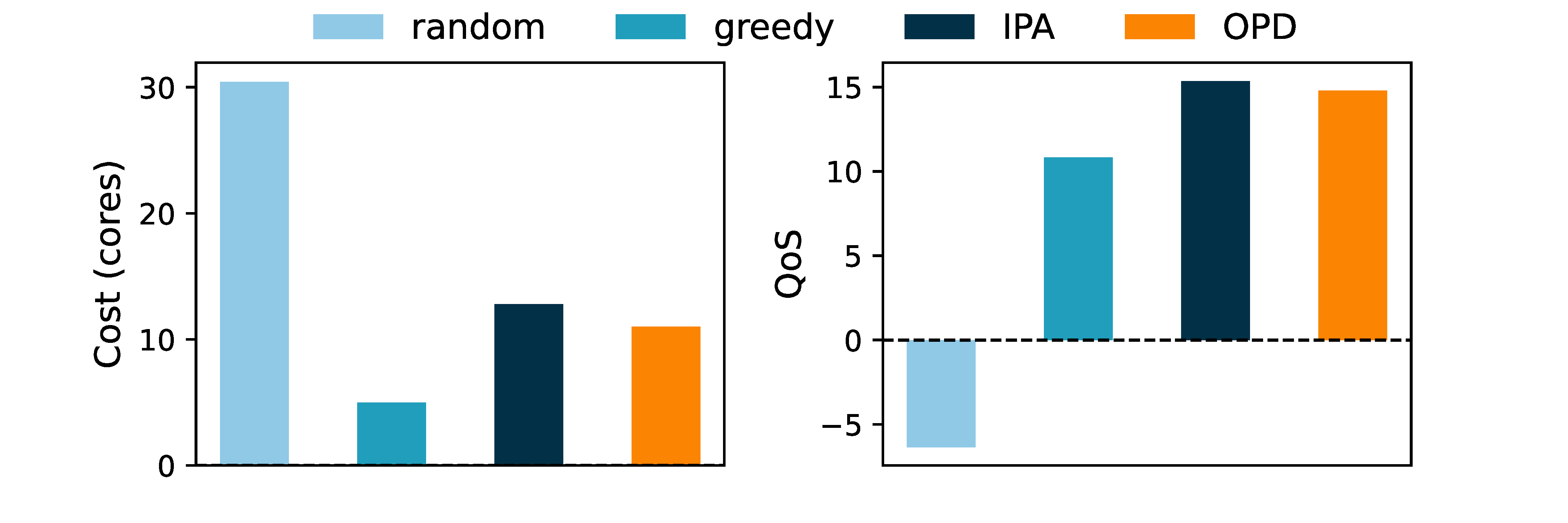}
        \caption{steady low workload}
        \label{fig: avg_low}
    \end{subfigure}
    \begin{subfigure}[!h]{0.5\textwidth}
        \centering
        \includegraphics[width=\textwidth]{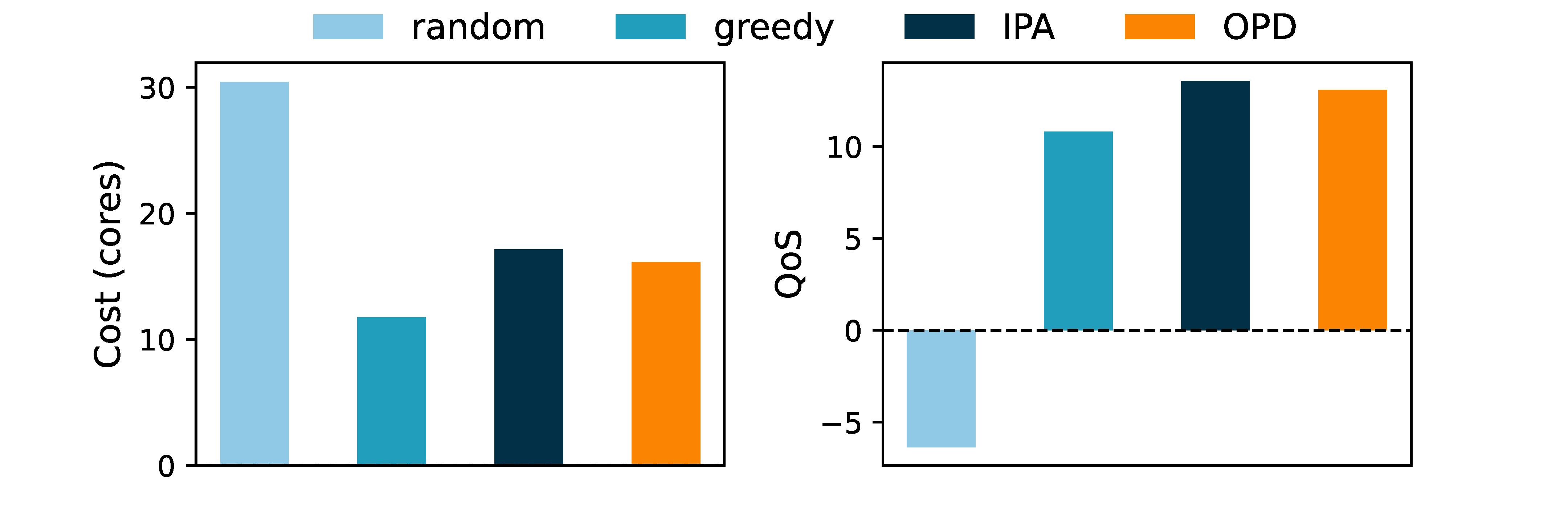}
        \caption{fluctuating workload}
        \label{fig: avg_flu}
    \end{subfigure}
    \begin{subfigure}[!h]{0.5\textwidth}
        \centering
        \includegraphics[width=\textwidth]{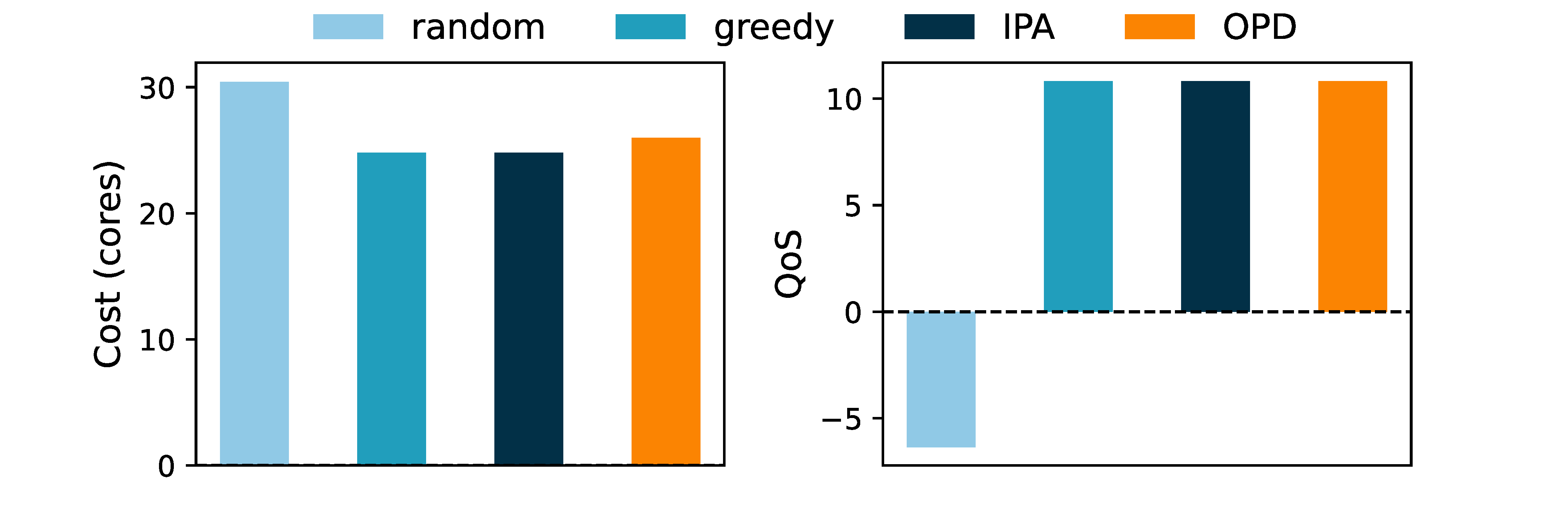}
        \caption{steady high workload}
        \label{fig: avg_high}
    \end{subfigure}
    \caption{Performance analysis under different workloads.}
    \label{fig: avg}
\end{figure}

\subsection{Experimental Results}

We evaluate the performance of the proposed OPD algorithm through a series of experiments in the Kubernetes cluster, comparing it with baseline algorithms.

\textbf{Performance with different workloads:} To minimize overhead from frequent switching, we established a 10-second adaptation interval and meticulously recorded performance data during each 1200-second operational cycle to evaluate four algorithms under various workloads. To ensure reproducibility and consistency, we fix the seed for all random generators.

\textit{Steady low load}: Fig. \ref{fig: time}(\subref{fig: time_low}) shows the performance of various algorithms under steady low load, offering a detailed comparison. The analysis indicates that all algorithms, except the random one, maintain consistent costs and QoS in this environment. While the greedy algorithm has the lowest cost, its QoS is relatively poor. In contrast, the IPA algorithm, despite being the most expensive, delivers the highest QoS. Our OPD algorithm strikes a balance between cost and QoS, positioning itself between greedy and IPA, and nearly matching IPA in service performance.

\textit{Fluctuating load}: Fig. \ref{fig: time}(\subref{fig: time_flu}) shows the algorithms' performance under fluctuating loads. Like in stable low-load conditions, the greedy algorithm excels in cost but fails to manage QoS during load fluctuations. The IPA algorithm offers stable services but is limited by its high cost in cost-sensitive settings. In contrast, the OPD algorithm effectively balances QoS and cost under varying loads, showcasing its adaptability.

\textit{Steady high load}: Fig. \ref{fig: time}(\subref{fig: time_high}) illustrates performance under steady high load conditions. In these environments, the high volume of task requests leads to increased costs for all algorithms. The random algorithm demonstrates significant cost fluctuations, reflecting instability in its resource allocation. In contrast, the greedy algorithm, IPA, and OPD maintain similar cost control and nearly identical QoS, suggesting they effectively balance cost and QoS to meet task demands.

Figs. \ref{fig: avg}(\subref{fig: avg_low}), \ref{fig: avg}(\subref{fig: avg_flu}), and \ref{fig: avg}(\subref{fig: avg_high}) show the average cost and QoS of various algorithms across a single workload cycle. Under steady low-load conditions, the OPD algorithm has a cost 120\% higher than the greedy algorithm but improves QoS by 36\%. Compared to the IPA method, it reduces costs by 16\% with only a minor 3.8\% drop in QoS. In fluctuating load conditions, OPD's cost is 37\% higher than the greedy algorithm's, yet it enhances QoS by 21\%. Relative to IPA, it decreases costs by 6\% while slightly lowering QoS by 3\%. In stable high-load scenarios, the greedy, IPA, and OPD algorithms exhibit nearly identical costs and QoS due to the high volume of requests, demonstrating OPD's effectiveness in balancing costs and QoS across different load conditions.

In short, the random algorithm exhibits significant cost and QoS fluctuations under different loads, indicating instability. The greedy algorithm is cost-effective but requires improved QoS. In contrast, both the IPA and OPD algorithms offer stable QoS. However, the IPA overlooks resource constraints and requires quicker decision-making. The OPD algorithm addresses these issues, optimally balancing cost and QoS while outperforming all baseline algorithms.

\begin{figure}[]
 \centering
 \includegraphics[width=1\linewidth]{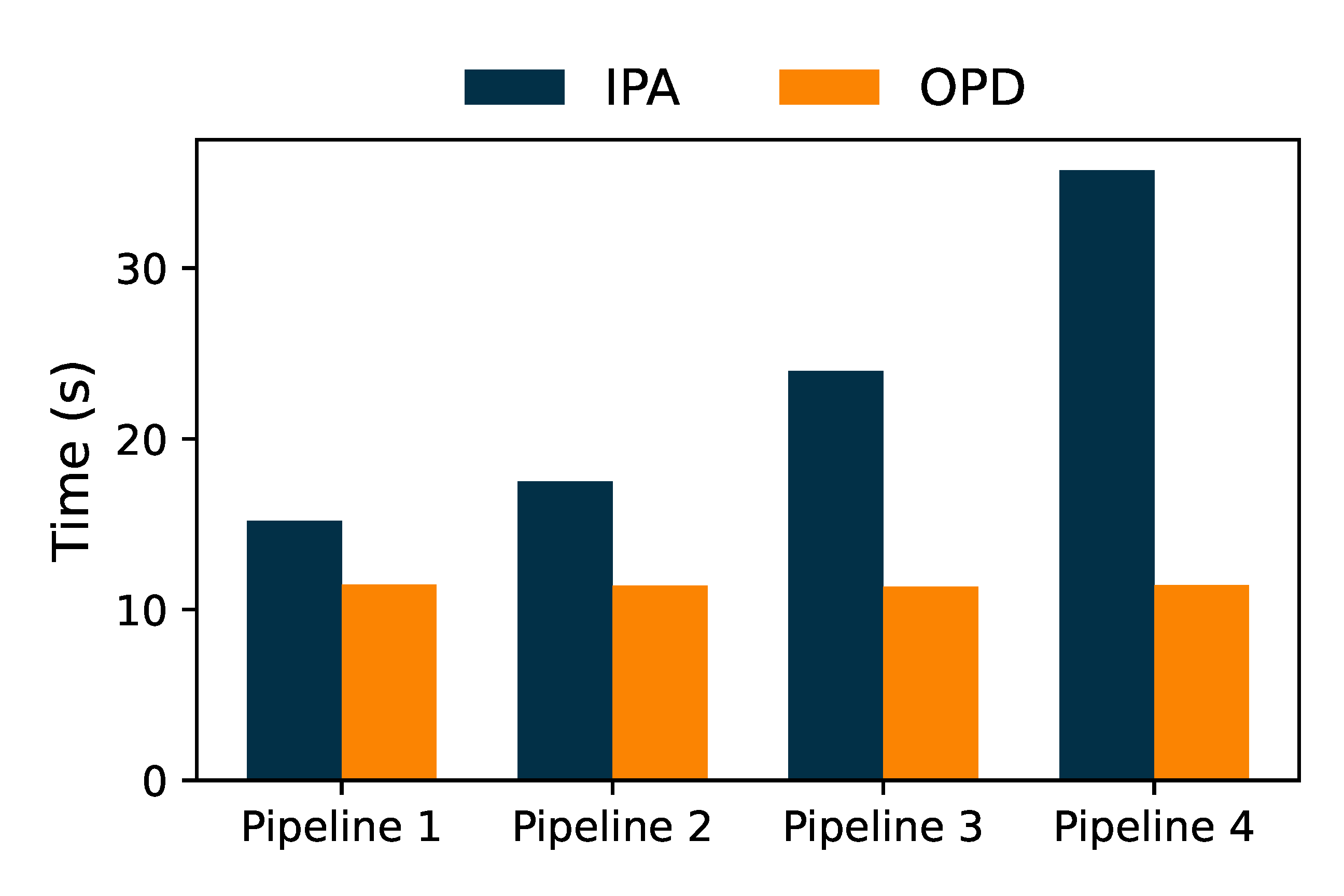}
 \centering
 \caption{Different pipelines decision time.}
 \label{fig: decision_time}
\end{figure}

\textbf{Performance with different pipelines:} To evaluate decision-making times across pipelines of varying complexities, we analyze four distinct pipelines with differing stages and model variants. Given the limited value of comparisons with random and greedy algorithms, we focus on the IPA and OPD methods. As shown in Fig. \ref{fig: decision_time}, IPA's decision-making time increases with pipeline complexity, while OPD's remains steady. As the number of tasks and model variants per stage increases, OPD's decision-making efficiency significantly surpasses IPA, with improvements of 32.5\%, 53.5\%, 111.6\%, and 212.8\% in processing a single workload cycle. Thus, the OPD algorithm effectively reduces decision-making time in complex multi-model inference pipelines, demonstrating the efficient applicability of the algorithm in large-scale and complex scenarios.

\begin{figure*}[htbp]
    \centering
    \includegraphics[width=\textwidth]{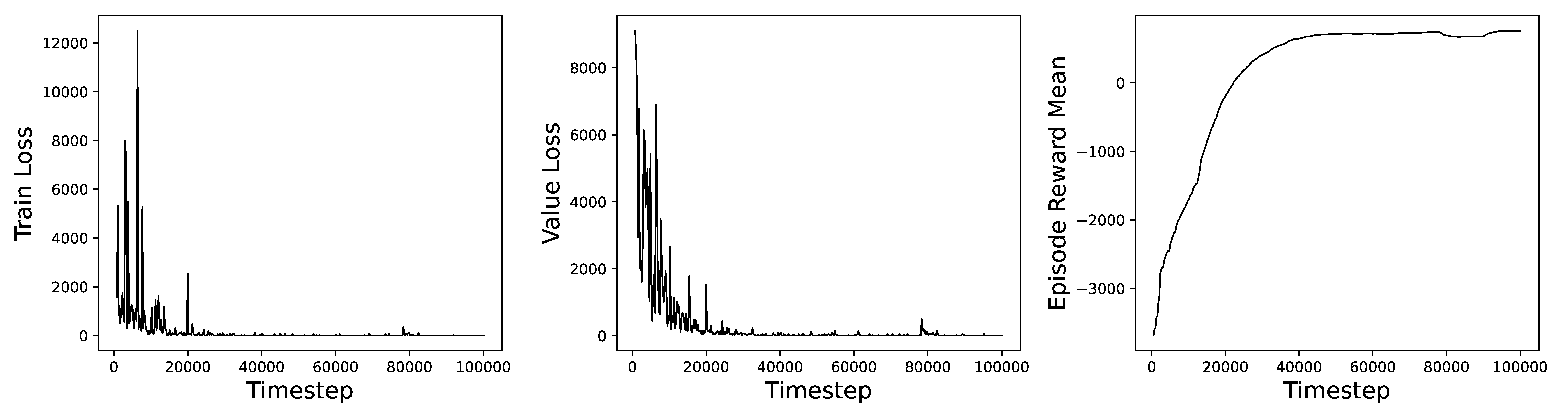}
    \caption{Training Loss, Value Loss, and Reward of the OPD algorithm.}
    \label{fig: rl}
\end{figure*}

\textbf{Security of the OPD algorithm:} In a distributed environment, dynamically adjusting the configurations of a multi-model inference pipeline may introduce potential security risks. To prevent these risks, we set restrictions such as the number of replicas and batch sizes in the OPD algorithm to prevent malicious users or attackers from manipulating the algorithm to select inappropriate configurations, thereby avoiding problems such as system overload and service interruption.

\textbf{Convergence of the OPD algorithm:} To illustrate the convergence of the OPD algorithm, Fig. \ref{fig: rl} displays the training loss, value loss, and mean episode reward. As training progresses, both the training loss and value function loss decrease rapidly, eventually stabilizing around a specific value, indicating convergence. The figure also shows that the OPD algorithm's reward converges to a higher value. Overall, the OPD algorithm exhibits a remarkably rapid convergence rate, demonstrating its ability to learn optimal policies in a short timeframe.

\section{Conclusion}
\label{section: conclusion}

We developed an online configuration decision algorithm for multi-model inference pipelines. First, we thoroughly modeled the OPD problem, taking into account cost, QoS, and device resource limitations. Next, we introduced a feature extraction method using a residual network to effectively capture node and pipeline state features. Finally, we proposed an OPD algorithm with an LSTM predictor based on policy gradient reinforcement learning for making configuration decisions. Experiments conducted on a real Kubernetes cluster showed that our algorithm outperforms baseline methods. Future work will focus on optimizing GPU resource allocation for multi-model inference pipelines in the Kubernetes cluster to enhance inference speed, reduce latency, and achieve higher QoS.

\bibliographystyle{ieeetr}
\small\bibliography{reference}

\vspace{12pt}
\color{red}

\end{document}